\documentclass[10pt, twocolumn]{iopart}

\usepackage{graphicx}
\usepackage{dcolumn}
\usepackage{bm}
\usepackage{siunitx}
\usepackage{xcolor}

\usepackage[style=nature,backend=biber,url=false,]{biblatex}
\bibliography{CDI_references2}

\begin{document}
\title[Non-invasive Current Density Imaging of Lithium-Ion Batteries]{Non-invasive Current Density Imaging of Lithium-Ion Batteries}
\author{Mark G. Bason$^1$, Thomas Coussens$^1$, Matthew Withers$^2$, Christopher Abel$^1$, Gary Kendall$^2$, Peter Kr\"uger$^1$}
\address{$^1$ Department of Physics and Astronomy, University of Sussex. }
\address{$^2$ CDO2 Ltd.}

\begin{abstract}
The rapid pace of replacing fossil fuel propelled transport by electric vehicles is critically dependent on high-performing, high energy density and efficient batteries. Optimal and safe use of existing battery cells and development of much-needed novel battery chemistries and geometries require a large range of diagnostic and monitoring tools. While structural and chemical information is readily extracted through a host of imaging techniques, non-invasive functional detection of interior battery processes remains limited. Here we introduce sensitive magnetometry performed outside the battery that reveals a battery cell's internal current distribution. As a key application, we use an array of sensors to image the magnetic field present under cycling of a pouch cell between charge states. We find good agreement between measured and modelled fields with sufficient resolution to detect percent-level deviations around areas of high current density. This opens the path towards rapid and reliable assessment throughout the battery life cycle, from battery development and  manufacturing quality assurance to operational safety and optimised use.
\end{abstract}

\maketitle
\ioptwocol
As the adoption of electrochemical devices for energy storage continues to grow, the pursuit of higher energy density, power density, increased safety and reduced costs continues apace \cite{Passerini2016a}. 
The application-specific demands placed upon these devices are sufficiently diverse to mean no single technology addresses all requirements. 
Hence, strategies for improving electric batteries, super-capacitors and fuel cells are currently being pursued.
In battery research, avenues of investigation include the identification of new cathode materials \cite{Ceder1998,Fergus2010a}, solid electrolytes \cite{Zhao2020a}, and transport processes in solid-state batteries \cite{Yu2017}. In addition, optimising design and developing monitoring techniques and failure-mitigation strategies, such as battery management systems~\cite{Xiong2018a} are helping to improve the efficiency of existing technology.
A number of diagnostic and characterisation techniques exist to aid in these developments \cite{Cabana2014}.
These can be broadly separated into those that focus on micro- and nano-structural investigations and macro or bulk measurements. 

 The ability to characterise materials on multiple length scales is driven by the need to study diverse physical processes \cite{Shearing2012a}, as no single technique is able to capture all the phenomena associated with electrochemical devices. 
 The former group of functional and structural analysis techniques includes x-ray, NMR \cite{Klamor2015}, optical microscopy, TEM \cite{Yuan2017} and neutron depth profiling \cite{Han2019}.
 The latter typically focuses on the measurement of bulk-quantities, such as cell voltage, current, temperature and impedance \cite{Barai2019}. 
 Through the measurements of these parameters using appropriate techniques, such as EIS \cite{Barsoukov2018}, Open-circuit voltages~\cite{Xing2014} and thermometry \cite{Murashko2013,Yazdanpour2014,Raijmakers2019},  can provide estimators of state of charge (SoC), state of health (SoH) and capacity.

Of particular importance to these measurement techniques is that they are non-invasive, relatively inexpensive and track activity in real time.  Electrochemical acoustic techniques have recently been successfully used to relate changes in sound speed to SoC and SoH \cite{Hsieh2015}. Structural changes have been detected through magnetic susceptibility in lithium-ion cells \cite{Hu2020}. While in polymer electrolyte membrane fuel cells, magnetic tomography~\cite{Wieser2000} has been used to monitor current density changes for fault detection~\cite{Ifrek2019}.
On the microscale, advanced imaging of small quantities of cells using NMR are also possible \cite{Ilott2018}. 
X-ray tomography and thermal imaging can also track thermal and structural evolution of a cell during thermal runaway \cite{Finegan2015}. 
However, the expense of these techniques tend to prohibit testings of large numbers of cells.

In contrast, more intrusive approaches can also be used to characterise cells.
Examples of these include the fabrication of cells with in-built sensors \cite{Fleming2019} and the measurement of individual electrode potentials using multi-tab approaches \cite{Erhard2017} or multiple electrodes \cite{Ng2009}. 
    \begin{figure}[h]
    \includegraphics[]{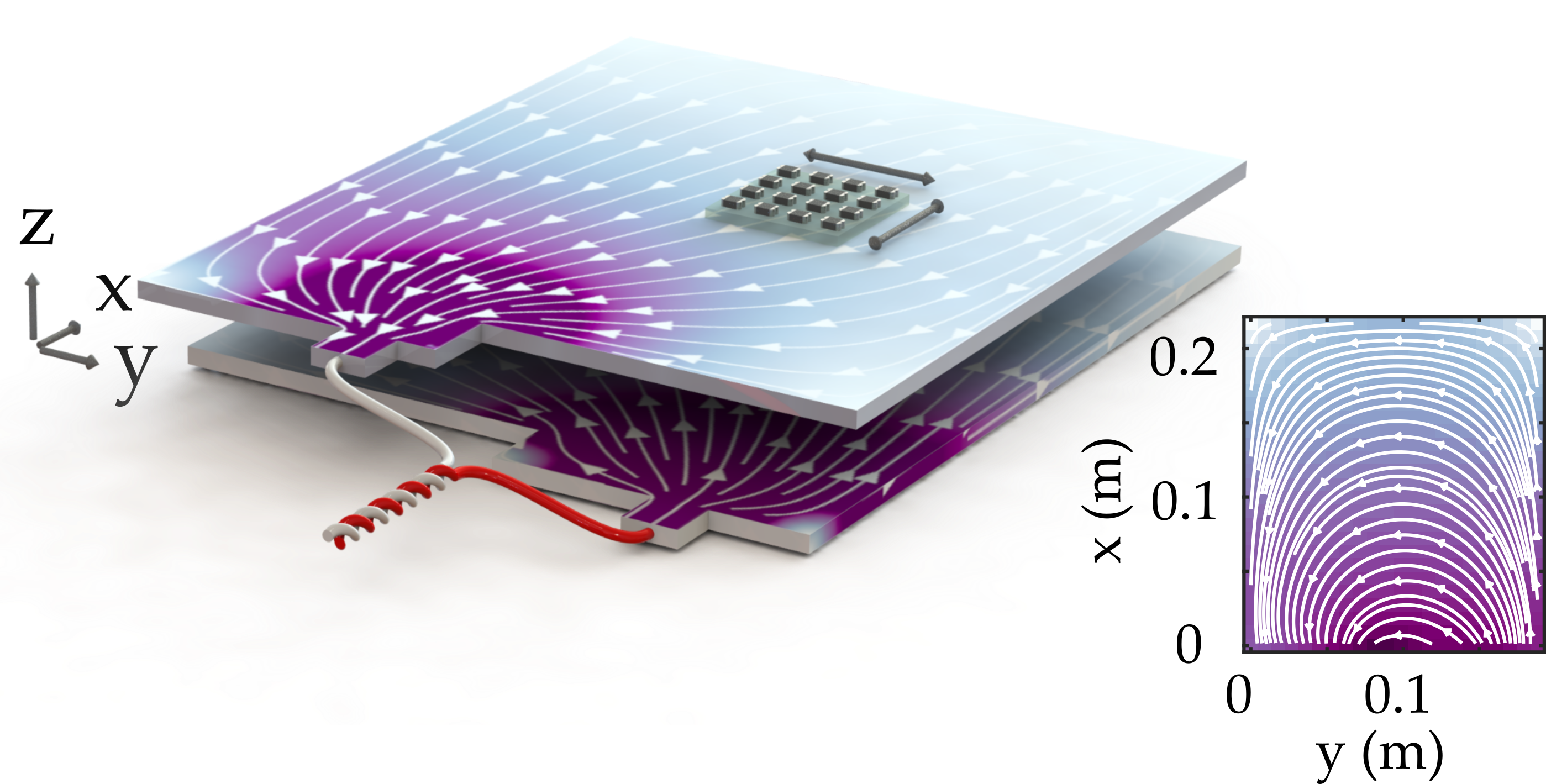}
    \caption{\label{fig:1} The current flow within a Li-ion battery and magnetic field gives rise to a magnetic field which is measured by an magnetometer array. The image shows a combination of current collectors and electrodes while the separator is suppressed for clarity.}
    \end{figure}
 
Here we introduce a direct, non-invasive measurement of electrochemical activity inside cells, which will give entirely new insights into the performance and safety of battery cells during research and development, facilitate manufacture and quality control, and facilitate optimal and safe operation. Our technique exploits the distribution of magnetic fields around a lithium-ion cell under load to perform a non-invasive, in-operando and contact-free measurement of local current densities. These current flows arise as a result of a combination of overpotentials and impedance of an electrochemical cell and are typically described by the Newman model of porous electrodes~\cite{Newman1962}.
 We focus on `active' cell behaviour, i.e. the movement of charge during cell cycling.
     Of fundamental interest to understanding battery behaviour, current density is critical in causing SoC inhomogenities, predicting heat generation, SEI thickness and formation, inhomogeneous extraction of lithium-ions \cite{Kindermann2017}, and lithium-plating \cite{Sieg2019a,Colclasure2019}. 
     Ultimately, a deeper understanding of these effects will help to maximise battery lifetime performance.
     In particular, in-homogeneous current density distributions are important in understanding lithium-ion transport mechanisms, including long-term equilibration processes \cite{Kindermann2017}. Direct measurements of current density should also complement modelling and simulation of battery behaviour during constant-current discharge processes \cite{Taheri2014,Meyer2013}.\\ 
   
     Note that, in contrast to thermal imaging, near-field measurements of magnetic fields are instantaneous - any movement of electrical charge causes an immediate change in the surrounding magnetic field. 
     
     We focus on lithium-ion batteries, in particular on pouch cells, for which the task of computing 2D current density images from magnetic field maps is simpler than for other cell formats, such as cylindrical and prismatic cells.
      We expect this method to have usage in cell design, characterisation and diagnosis.
      The ability of spatially resolving regions of high electrical conductivities will allow the monitoring of dendrite growth in real-time, for example in lithium-metal batteries \cite{Varzi2016}. In addition, it is expected to be useful for the investigation of conduction phenomena \cite{Park2010} and the characterisation of battery chemistries for which open-circuit voltage is not an accurate measure of a cell's SoC, such as lithium-sulphur \cite{Manthiram2014}.

\section{Experimental}

     An overview of the system can be seen in Fig.\,\ref{fig:1}. 
     A 2$^{\textrm{nd}}$ generation AESC pouch battery, with a rated capacity of \SI{66}{\ampere\hour}, is disassembled to extract a single cell \cite{Kovachev2019}.
     Each cell is composed of 35 electrode pairs, in addition to aluminium and copper current collectors, the dimensions of which are \SI{22}{\centi\metre}~$\times$ ~\SI{26}{\centi\metre}.\\
     The battery is situated \SI{6}{\milli\metre} below a 4~$\times$~4 fluxgate array (FGA) of single-axis magnetometers (Texas instruments DRV425); we note that other inexpensive magnetometers are also suitable~\cite{Huang2018}. 
     Each individual magnetometer is configured to record a component of the magnetic flux density, either $B_x$ and $B_y$, in a plane above the cell.
     To shield the sensor array from external magnetic noise, it is placed inside two layers of \SI{1}{\milli\metre} thick, high-permeability mu-metal which channel magnetic flux through its walls and acts to reduce the ambient magnetic flux density in the volume occupied by the FGA and battery.
     A motorised three-axis translation stage, with positioning reproducibility of around 10 microns is used to move the cell in a two-dimensional plane beneath the array \cite{Han2017}.\\
     The magnetic field is recorded in three states.
     Firstly, a battery cycler (Analog Devices AD8452) acting as a constant current source/sink, is used to charge or discharge the cell, providing a relative stability on the 10$^{-5}$ level.
     Secondly, the background magnetic field of~\SI{10}{\micro\tesla} is recorded with the cell disconnected from the battery cycler and positioned away from the FGA. This measurement is used to track the offset drift of the FGA, which is seen to non-trivially depend on temperature.  
     Thirdly, the passive field, believed to be due to nickel-plating on the copper current collector is measured. This is seen to be stable to charging and discharging to the level of \SI{100}{\nano\tesla}.
     The cell is not actively temperature stabilised, but rests at room temperature~\SI{21.5}{\degreeCelsius}.
         The choice of charging rate is a trade-off between capturing an narrow  SoC `window' versus the time taken to translate the FGA array over a region of interest. Each measurement takes~\SI{12}{minutes} to scan an area of around~\SI{150}{\centi\metre\squared} with a resolution of ~\SI{5}{\milli\metre}.
    \begin{figure}[h]
        \includegraphics[]{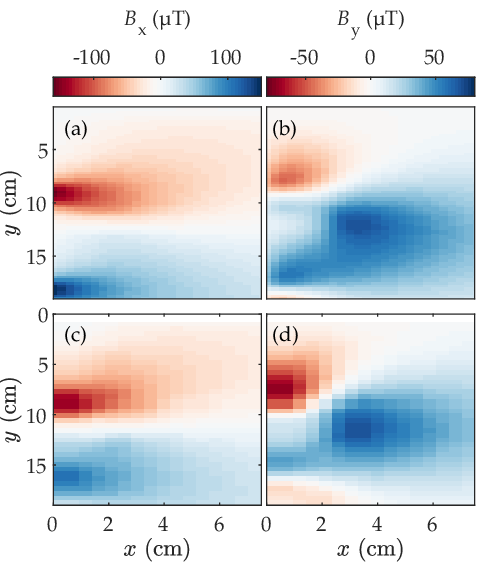}
        \caption{\label{fig:2} Magnetic fields generated by a~\SI{10}{\ampere} (0.2 C$_{\textrm{rate}}$) charge: (a) and (b) correspond to the $B_y$ and $B_x$ magnetic field components produced by the FEM predictions, respectively. The results of the measurements after background subtraction are shown in (c) and (d). The standard error associated with each measured field component measurement is around~\SI{100}{\nano\tesla}.}
    \end{figure}

\section{Simulation}

     To benchmark the accuracy of the magnetic field measurements, we compare our magnetic field images to those predicted by a finite-element model.
     This model, based on a single pair of electrodes, is used to calculate the stationary solutions of Poisson's equation for the voltage of the cell. Conservation of the current density $J$ is maintained through the cell, by ensuring $\nabla\cdot{}J = 0$, except at the boundaries needed for the terminals. At these points, the boundary conditions are $\nabla\cdot{}J = Q_{j,v}$.
     $Q_{j,v}$ corresponds to a distributed current source used as the input to the battery, the value of which is set to ensure the specified current flows through the battery.
    The model is solved for an intermediate layer of thickness \SI{150}{\micro\metre}, which represents both positive and negative porous electrodes and separator. 
     The effective conductivity, a combination of electrical and ionic conductivity \cite{Taheri2014}, is set to \SI{1e-3}{\siemens\per\metre}, which corresponds to the value necessary to provide a potential drop in the open-circuit voltage.
     The $z$ component of the current density thus combines intercalation of lithium ions in active materials on both sides of the electrodes.
     \\
     Using 11824 mesh points, with a density of around \SI{3}{\per\centi\metre\squared}, the model is solved to provide magnetic field distributions on the millimetre scale.
     The current density, $\bm{J}$, at each point in the geometry is then used to calculate the magnetic field at a distance $d$ above the cell using Biot-Savart's law. 
    \begin{equation}
        \bm{B(r)} = \frac{\mu_0{}d}{4\pi}\int\frac{\bm{J}(\bm{r'})\times(\bm{r - r'})}{\mid{\bm{r-r'}}\mid^3}d^3\bm{r'},
    \end{equation}
     where $\bm{r}$ is the position in space at which the magnetic field is calculated, $\bm{r'}$ is the position of the current density and $\mu_0{}$ is the permeability of free space. 
  
\section{Results and Discussion}

\begin{figure}[h]
    \includegraphics[width=80mm]{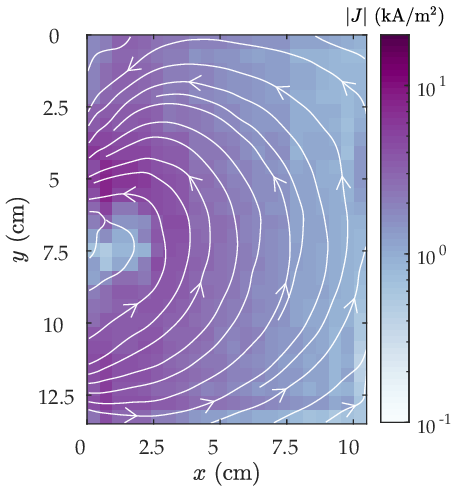}
    \caption{\label{fig:3} Reconstructed current density image corresponding to a single electrode pair, during a \SI{10}{\ampere} discharge. The top half of the image is shown here, with the battery tabs located at $x=0$.}
    \end{figure}

     An example of the two magnetic fields after background subtraction is shown in Fig.\,\ref{fig:2}.
     Magnetic fields of over \SI{100}{\micro\tesla} are recorded, in good quantitative agreement with the FEM simulations. The strength of the magnetic field is dominated by the high current density at the input and output of the current tabs - a reflection on the particular pouch cell geometry under investigation.
     As the current enters the cell along the $x$ direction, it creates a large, positive field in the $y$ direction. 
     After the current starts to spread along the length of the battery, a $y$ component of the current density is established with a corresponding magnetic field in the $x$ direction.
These images are used to calculate the current density using an approach based on Green's functions~\cite{Roth1989}. 
    Making use of the convolution theorem, a current restricted to 2D can be expressed as the spatially-filtered Fourier transform of the magnetic field. For the y-component of the current density, this is
    \begin{equation}
        j_y(x,y) = -\frac{2}{\mu_0{}d}\frac{k_y}{k_x}e^{\sqrt{k_x^2+k_y^2}z}b_x(k_x,k_y,z),
    \end{equation}
    where $k_{x,y}=2\pi/d$ and $b_z$ is the Fourier transform of $B_x$.
    Likewise, a similar expression yields $j_x$ in terms of the scaled Fourier transform of $B_y$~\cite{Roth1989}.
     Thus, an accurate translation of magnetic field to current density requires precise knowledge of the sensor-cell separation and cell thickness.
     Using such an approach, the battery is treated as a 2D object. As the electrode thickness in pouch cells are typically orders of magnitude smaller than their length and width, this is a good approximation.
     Thus, the vector sum of each current component in 3D space is used to produce a single vector at each position in 2D space. Note that the combination of measurement of the out-of-plane magnetic field component and measurement at different heights should, in principle, be used to develop 3D images.
    A typical current density image, taken during cell discharge, is shown in Fig.\,\ref{fig:3}.  
     As the current density norm shows, the current is concentrated at the input and dies off away from these tabs. Since the exact current density distribution is characteristic of the cell's effective conductivity and cell geometry, any changes in the magnetic field distribution could be used to infer cell properties, such as lithiation state and diffusion coefficients.   
\begin{figure}[h]
    \includegraphics[]{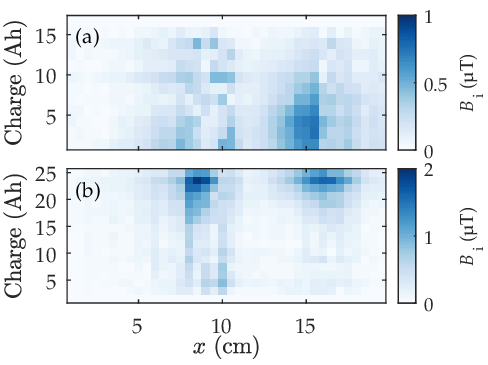}
    \caption{\label{fig:4} Change in total magnetic field around the battery terminals during a period of discharging (a) and charging (b) at \SI{10}{\ampere}.}
\end{figure}
To investigate effects not captured by our model, such as changes with SoC, we now measure magnetic fields changes that emerge at different charge stages. Starting with a battery with an open cell potential of around \SI{3.7}{\volt}, we cycle the battery between charge states with a constant current of \SI{10}{\ampere} in stages of \SI{1.3}{\ampere\hour}.
     We look for magnetic field changes between each individual image and that corresponding to the first SoC. Based upon these observations, these could then be used to refine future models.
     While it is possible to convert these recordings into full current density images, doing so shows that the main changes in current density occurs around the tabs. Thus, Fig.\,\ref{fig:4} shows the magnetic field differences for the first column of readings, i.e. those closest to the current collectors. In  Fig.\,\ref{fig:4}(a), we show the differences in total magnetic field components, $\sqrt{\left(\Delta{}B_x^2 + \Delta{}B_y^2\right)}$, corresponding to the difference between an initially charged battery and subsequent charge states. Likewise, in Fig.\,\ref{fig:4}(b) the difference from an initially empty battery is shown.
     While the changes correspond to a small fraction of the total magnetic field - around 1\%, they can be clearly identified above the noise.  Note that the changes around the respective battery tabs are not symmetric, but emerge at different SoC.
     The changes in the two field components do, to a large extent, reflect the nature of the fields shown in Fig.\,\ref{fig:1}. As the changes are around 100 times larger than the noise level of the battery cycler, changes in the charging current are unable to explain the observed changes. Nor is the change in effective conductivity associated with different lithiation states expected to be large enough to account for this effect. Likewise, relative movements between the FGA and the pouch cell would produce magnetic field changes that are inconsistent with those detected.
     It is possible that strain-induced effects~\cite{Zhang2018a}, for example, through isotropic expansion of negative electrodes may contribute~\cite{Timmons2007a}, or localised heating may account for the changes.
     Simultaneous measurements of strain, temperature and magnetic field are beyond the scope of this paper, but are good candidates for subsequent studies.
     Likewise, the development of larger magnetometer arrays will allow the simultaneous capture of data from across the battery and remove the need to translate the FGA. This will allow the capture of current density images at higher rates of charge/discharge - situations in which equivalent circuit models often struggle to describe battery behaviour.\\
     Given the ability to detect magnetic field changes with sub-femtotesla sensitivity \cite{Kominis2003} and the development of sensor arrays~\cite{Borna2019a}, we envision current density images with over six orders of magnitude lower noise-floors that those presented here. These will allow measurements of current densities on the \SI{1}{\nano\ampere\centi\metre\squared} scale. This opens the possibility for self-discharge measurements of disconnected batteries to assess cell conditions after manufacture or as part of maintenance measurements.
\section{Conclusion}
     We have shown how highly-sensitive magnetometer arrays can be used to non-invasively generate in-situ current density images of lithium-ion pouch cells under load. We confirm the general validity of FEM models while also identifying localised, SOC-dependent deviations on the percent level. In the near-term, we expect that the method will be used to detect current density `hot-spots' in real-time, for example those associated with dendrite formation or mechanical integrity issues such as soft and hard short circuits. In addition, the technique will also aid the verification of new electrochemical cell designs in which the homogeneity of current density is critical. Introducing high-sensitivity and quantum magnetometers for functional battery imaging will have large impacts in research and development; manufacture; safe-use and recycling due to its speed, non-invasive nature and highly-resolved functional mapping. 
\section{Acknowledgments}
The authors would like to acknowledge support from Innovate UK (project numbers 133724 and 42186). We would like to thank D. Shiers and R. Puddy for their technical assistance and E. Eweka for insightful discussions.


\printbibliography

@article{Klamor2015,
    title = {{7Li in situ 1D NMR imaging of a lithium ion battery}},
    year = {2015},
    journal = {Physical Chemistry Chemical Physics},
    author = {Klamor, S. and Zick, K. and Oerther, T. and Schappacher, F. M. and Winter, M. and Brunklaus, G.},
    number = {6},
    month = {1},
    pages = {4458--4465},
    volume = {17},
    publisher = {Royal Society of Chemistry},
    url = {http://xlink.rsc.org/?DOI=C4CP05021E},
    doi = {10.1039/c4cp05021e},
    issn = {14639076}
}

@article{Barai2019,
    title = {{A comparison of methodologies for the non-invasive characterisation of commercial Li-ion cells}},
    year = {2019},
    journal = {Progress in Energy and Combustion Science},
    author = {Barai, Anup and Uddin, Kotub and Dubarry, Matthieu and Somerville, Limhi and McGordon, Andrew and Jennings, Paul and Bloom, Ira},
    month = {5},
    pages = {1--31},
    volume = {72},
    publisher = {Pergamon},
    url = {https://www.sciencedirect.com/science/article/pii/S0360128518300996},
    doi = {10.1016/j.pecs.2019.01.001},
    issn = {03601285}
}

@article{Yazdanpour2014,
    title = {{A Distributed Analytical Electro-Thermal Model for Pouch-Type Lithium-Ion Batteries}},
    year = {2014},
    journal = {Journal of The Electrochemical Society},
    author = {Yazdanpour, Maryam and Taheri, Peyman and Mansouri, Abraham and Bahrami, Majid},
    number = {14},
    month = {9},
    pages = {A1953-A1963},
    volume = {161},
    publisher = {The Electrochemical Society},
    url = {http://jes.ecsdl.org/lookup/doi/10.1149/2.1191412jes},
    doi = {10.1149/2.1191412jes},
    issn = {0013-4651}
}

@article{Ng2009,
    title = {{A multiple working electrode for electrochemical cells: A tool for current density distribution studies}},
    year = {2009},
    journal = {Angewandte Chemie - International Edition},
    author = {Ng, See How and La Mantia, Fabio and Nov{\'{a}}k, Petr},
    number = {3},
    month = {1},
    pages = {528--532},
    volume = {48},
    publisher = {John Wiley {\&} Sons, Ltd},
    url = {http://doi.wiley.com/10.1002/anie.200803981},
    doi = {10.1002/anie.200803981},
    issn = {14337851}
}

@article{Park2010,
    title = {{A review of conduction phenomena in Li-ion batteries}},
    year = {2010},
    journal = {Journal of Power Sources},
    author = {Park, Myounggu and Zhang, Xiangchun and Chung, Myoungdo and Less, Gregory B. and Sastry, Ann Marie},
    number = {24},
    month = {12},
    pages = {7904--7929},
    volume = {195},
    publisher = {Elsevier},
    url = {https://www.sciencedirect.com/science/article/pii/S0378775310010463},
    doi = {10.1016/J.JPOWSOUR.2010.06.060},
    issn = {0378-7753}
}

@article{Raijmakers2019,
    title = {{A review on various temperature-indication methods for Li-ion batteries}},
    year = {2019},
    journal = {Applied Energy},
    author = {Raijmakers, L. H.J. and Danilov, D. L. and Eichel, R. A. and Notten, P. H.L.},
    month = {4},
    pages = {918--945},
    volume = {240},
    publisher = {Elsevier},
    url = {https://www.sciencedirect.com/science/article/pii/S0306261919303757?via%3Dihub},
    doi = {10.1016/j.apenergy.2019.02.078},
    issn = {03062619}
}

@article{Kominis2003,
    title = {{A subfemtotesla multichannel atomic magnetometer}},
    year = {2003},
    journal = {Nature},
    author = {Kominis, I. K. and Kornack, T. W. and Allred, J. C. and Romalis, M. V.},
    number = {6932},
    pages = {596--599},
    volume = {422},
    url = {http://dx.doi.org/10.1038/nature01484%5Cnhttp://www.nature.com/nature/journal/v422/n6932/abs/nature01484.html%5Cnhttp://www.nature.com/nature/journal/v422/n6932/pdf/nature01484.pdf},
    isbn = {0028-0836 (Print) 0028-0836 (Linking)},
    doi = {10.1038/nature01484},
    issn = {0028-0836},
    pmid = {12686995}
}

@article{Yu2017,
    title = {{Accessing the bottleneck in all-solid state batteries, lithium-ion transport over the solid-electrolyte-electrode interface}},
    year = {2017},
    journal = {Nature Communications},
    author = {Yu, Chuang and Ganapathy, Swapna and Eck, Ernst R.H.Van and Wang, Heng and Basak, Shibabrata and Li, Zhaolong and Wagemaker, Marnix},
    number = {1},
    month = {12},
    pages = {1086},
    volume = {8},
    publisher = {Nature Publishing Group},
    url = {http://www.nature.com/articles/s41467-017-01187-y},
    doi = {10.1038/s41467-017-01187-y},
    issn = {20411723}
}

@article{Kovachev2019,
    title = {{Analytical dissection of an automotive Li-ion pouch cell}},
    year = {2019},
    journal = {Batteries},
    author = {Kovachev, Georgi and Schr{\"{o}}ttner, Hartmuth and Gstrein, Gregor and Aiello, Luigi and Hanzu, Ilie and Wilkening, H. Martin R. and Foitzik, Alexander and Wellm, Michael and Sinz, Wolfgang and Ellersdorfer, Christian},
    number = {4},
    month = {10},
    pages = {67},
    volume = {5},
    publisher = {Multidisciplinary Digital Publishing Institute},
    url = {https://www.mdpi.com/2313-0105/5/4/67},
    doi = {10.3390/batteries5040067},
    issn = {23130105}
}

@article{Varzi2016,
    title = {{Challenges and prospects of the role of solid electrolytes in the revitalization of lithium metal batteries}},
    year = {2016},
    journal = {Journal of Materials Chemistry A},
    author = {Varzi, Alberto and Raccichini, Rinaldo and Passerini, Stefano and Scrosati, Bruno},
    number = {44},
    month = {11},
    pages = {17251--17259},
    volume = {4},
    publisher = {The Royal Society of Chemistry},
    url = {http://xlink.rsc.org/?DOI=C6TA07384K},
    doi = {10.1039/c6ta07384k},
    issn = {20507496}
}

@inproceedings{Zhang2018a,
    title = {{Characterization of external pressure effects on lithium-ion pouch cell}},
    year = {2018},
    booktitle = {Proceedings of the IEEE International Conference on Industrial Technology},
    author = {Zhang, Yuan Ci and Briat, Olivier and Deletage, Jean Yves and Martin, Cyril and Gager, Guillaume and Vinassa, Jean Michel},
    month = {2},
    pages = {2055--2059},
    volume = {2018-Febru},
    publisher = {IEEE},
    url = {https://ieeexplore.ieee.org/document/8352505/},
    isbn = {9781509059492},
    doi = {10.1109/ICIT.2018.8352505}
}

@article{Zhao2020a,
    title = {{Designing solid-state electrolytes for safe, energy-dense batteries}},
    year = {2020},
    journal = {Nature Reviews Materials},
    author = {Zhao, Qing and Stalin, Sanjuna and Zhao, Chen Zi and Archer, Lynden A.},
    number = {3},
    month = {3},
    pages = {229--252},
    volume = {5},
    publisher = {Nature Publishing Group},
    url = {http://www.nature.com/articles/s41578-019-0165-5},
    doi = {10.1038/s41578-019-0165-5},
    issn = {20588437}
}

@article{Hsieh2015,
    title = {{Electrochemical-acoustic time of flight: In operando correlation of physical dynamics with battery charge and health}},
    year = {2015},
    journal = {Energy and Environmental Science},
    author = {Hsieh, A. G. and Bhadra, S. and Hertzberg, B. J. and Gjeltema, P. J. and Goy, A. and Fleischer, J. W. and Steingart, D. A.},
    number = {5},
    month = {5},
    pages = {1569--1577},
    volume = {8},
    publisher = {The Royal Society of Chemistry},
    url = {http://xlink.rsc.org/?DOI=C5EE00111K},
    doi = {10.1039/c5ee00111k},
    issn = {17545706}
}

@article{Sieg2019a,
    title = {{Fast charging of an electric vehicle lithium-ion battery at the limit of the lithium deposition process}},
    year = {2019},
    journal = {Journal of Power Sources},
    author = {Sieg, Johannes and Bandlow, Jochen and Mitsch, Tim and Dragicevic, Daniel and Materna, Torben and Spier, Bernd and Witzenhausen, Heiko and Ecker, Madeleine and Sauer, Dirk Uwe},
    month = {7},
    pages = {260--270},
    volume = {427},
    publisher = {Elsevier},
    url = {https://www.sciencedirect.com/science/article/abs/pii/S0378775319304586?via%3Dihub},
    doi = {10.1016/j.jpowsour.2019.04.047},
    issn = {03787753}
}

@article{Ifrek2019,
    title = {{Fault detection for polymer electrolyte membrane fuel cell stack by external magnetic field}},
    year = {2019},
    journal = {Electrochimica Acta},
    author = {Ifrek, Lyes and Rosini, Sébastien and Cauffet, Gilles and Chadebec, Olivier and Rouveyre, Luc and Bultel, Yann},
    month = {8},
    pages = {141--150},
    volume = {313},
    publisher = {Pergamon},
    url = {https://www.sciencedirect.com/science/article/pii/S0013468619309016},
    doi = {10.1016/j.electacta.2019.04.193},
    issn = {00134686}
}

@article{Han2019,
    title = {{High electronic conductivity as the origin of lithium dendrite formation within solid electrolytes}},
    year = {2019},
    journal = {Nature Energy},
    author = {Han, Fudong and Westover, Andrew S. and Yue, Jie and Fan, Xiulin and Wang, Fei and Chi, Miaofang and Leonard, Donovan N. and Dudney, Nancy J. and Wang, Howard and Wang, Chunsheng},
    month = {1},
    pages = {1},
    publisher = {Nature Publishing Group},
    url = {http://www.nature.com/articles/s41560-018-0312-z},
    doi = {10.1038/s41560-018-0312-z},
    issn = {2058-7546}
}

@article{Ceder1998,
    title = {{Identification of cathode materials for lithium batteries guided by first-principles calculations}},
    year = {1998},
    journal = {Nature},
    author = {Ceder, G. and Chiang, Y. M. and Sadoway, D. R. and Aydinol, M. K. and Jang, Y. I. and Huang, B.},
    number = {6677},
    month = {4},
    pages = {694--696},
    volume = {392},
    publisher = {Nature Publishing Group},
    url = {http://www.nature.com/articles/33647},
    doi = {10.1038/33647},
    issn = {00280836}
}

@article{Huang2018,
    title = {{Implementation of 16-Channel AMR Sensor Array for Quantitative Mapping of Two-Dimension Current Distribution}},
    year = {2018},
    journal = {IEEE Transactions on Magnetics},
    author = {Huang, Guan Wei and Jeng, Jen Tzong},
    number = {11},
    month = {11},
    pages = {1--5},
    volume = {54},
    url = {https://ieeexplore.ieee.org/document/8434124/},
    doi = {10.1109/TMAG.2018.2844290},
    issn = {00189464}
}

@article{Finegan2015,
    title = {{In-operando high-speed tomography of lithium-ion batteries during thermal runaway}},
    year = {2015},
    journal = {Nature Communications},
    author = {Finegan, Donal P. and Scheel, Mario and Robinson, James B. and Tjaden, Bernhard and Hunt, Ian and Mason, Thomas J. and Millichamp, Jason and Di Michiel, Marco and Offer, Gregory J. and Hinds, Gareth and Brett, Dan J.L. and Shearing, Paul R.},
    number = {1},
    month = {11},
    pages = {6924},
    volume = {6},
    publisher = {Nature Publishing Group},
    url = {http://www.nature.com/articles/ncomms7924},
    doi = {10.1038/ncomms7924},
    issn = {20411723}
}

@article{Timmons2007a,
    title = {{Isotropic Volume Expansion of Particles of Amorphous Metallic Alloys in Composite Negative Electrodes for Li-Ion Batteries}},
    year = {2007},
    journal = {Journal of The Electrochemical Society},
    author = {Timmons, A. and Dahn, J. R.},
    number = {5},
    month = {3},
    pages = {A444},
    volume = {154},
    publisher = {IOP Publishing},
    url = {https://iopscience.iop.org/article/10.1149/1.2711075},
    doi = {10.1149/1.2711075},
    issn = {00134651}
}

@article{Passerini2016a,
    title = {{Lithium and lithium-ion batteries: Challenges and prospects}},
    year = {2016},
    journal = {Electrochemical Society Interface},
    author = {Passerini, Stefano and Scrosati, Bruno},
    number = {3},
    month = {1},
    pages = {85--87},
    volume = {25},
    publisher = {IOP Publishing},
    url = {https://iopscience.iop.org/article/10.1149/2.F09163if},
    doi = {10.1149/2.F09163if},
    issn = {19448783}
}

@article{Borna2019a,
    title = {{Magnetic Source Imaging Using a Pulsed Optically Pumped Magnetometer Array}},
    year = {2019},
    journal = {IEEE Transactions on Instrumentation and Measurement},
    author = {Borna, Amir and Carter, Tony R. and Derego, Paul and James, Conrad D. and Schwindt, Peter D.D.},
    number = {2},
    month = {2},
    pages = {493--501},
    volume = {68},
    url = {https://ieeexplore.ieee.org/document/8418461/},
    doi = {10.1109/TIM.2018.2851458},
    issn = {00189456}
}

@article{Kindermann2017,
    title = {{Measurements of lithium-ion concentration equilibration processes inside graphite electrodes}},
    year = {2017},
    journal = {Journal of Power Sources},
    author = {Kindermann, Frank M. and Osswald, Patrick J. and Klink, Stefan and Ehlert, Günter and Schuster, Jörg and Noel, Andreas and Erhard, Simon V. and Schuhmann, Wolfgang and Jossen, Andreas},
    month = {2},
    pages = {638--643},
    volume = {342},
    publisher = {Elsevier},
    url = {https://www.sciencedirect.com/science/article/abs/pii/S0378775316317967},
    doi = {10.1016/j.jpowsour.2016.12.093},
    issn = {03787753}
}

@article{Shearing2012a,
    title = {{Multi Length Scale Microstructural Investigations of a Commercially Available Li-Ion Battery Electrode}},
    year = {2012},
    journal = {Journal of The Electrochemical Society},
    author = {Shearing, P. R. and Brandon, N. P. and Gelb, J. and Bradley, R. and Withers, P. J. and Marquis, A. J. and Cooper, S. and Harris, S. J.},
    number = {7},
    pages = {A1023-A1027},
    volume = {159},
    doi = {10.1149/2.053207jes},
    issn = {0013-4651}
}

@article{Wieser2000,
    title = {{New technique for two-dimensional current distribution measurements in electrochemical cells}},
    year = {2000},
    journal = {Journal of Applied Electrochemistry},
    author = {Wieser, C. and Helmbold, A. and G{\"{u}}lzow, E.},
    number = {7},
    pages = {803--807},
    volume = {30},
    publisher = {Springer},
    url = {http://link.springer.com/10.1023/A:1004047412066},
    doi = {10.1023/A:1004047412066},
    issn = {0021891X}
}

@article{Fergus2010a,
    title = {{Recent developments in cathode materials for lithium ion batteries}},
    year = {2010},
    journal = {Journal of Power Sources},
    author = {Fergus, Jeffrey W.},
    number = {4},
    month = {2},
    pages = {939--954},
    volume = {195},
    publisher = {Elsevier},
    url = {https://www.sciencedirect.com/science/article/pii/S0378775309015304?casa_token=9n2OIvxOKQEAAAAA:dR7oRI1aHgOfUhuaNEwJqB7TN1H29GKD6kc3w6hWwZBiBDBEvY0Y_jInoqIQVd2VbLnTiNz7CV0},
    doi = {10.1016/j.jpowsour.2009.08.089},
    issn = {03787753}
}

@article{Ilott2018,
    title = {{Rechargeable lithium-ion cell state of charge and defect detection by in-situ inside-out magnetic resonance imaging}},
    year = {2018},
    journal = {Nature Communications},
    author = {Ilott, Andrew J. and Mohammadi, Mohaddese and Schauerman, Christopher M. and Ganter, Matthew J. and Jerschow, Alexej},
    number = {1},
    month = {12},
    pages = {1776},
    volume = {9},
    publisher = {Nature Publishing Group},
    url = {http://www.nature.com/articles/s41467-018-04192-x},
    doi = {10.1038/s41467-018-04192-x},
    issn = {2041-1723}
}

@article{Colclasure2019,
    title = {{Requirements for Enabling Extreme Fast Charging of High Energy Density Li-Ion Cells while Avoiding Lithium Plating}},
    year = {2019},
    journal = {Journal of The Electrochemical Society},
    author = {Colclasure, Andrew M. and Dunlop, Alison R. and Trask, Stephen E. and Polzin, Bryant J. and Jansen, Andrew N. and Smith, Kandler},
    number = {8},
    month = {4},
    pages = {A1412-A1424},
    volume = {166},
    publisher = {IOP Publishing},
    url = {https://iopscience.iop.org/article/10.1149/2.0451908jes},
    doi = {10.1149/2.0451908jes},
    issn = {0013-4651}
}

@article{Hu2020,
    title = {{Sensitive magnetometry reveals inhomogeneities in charge storage and weak transient internal currents in Li-ion cells}},
    year = {2020},
    journal = {Proceedings of the National Academy of Sciences},
    author = {Hu, Yinan and Iwata, Geoffrey Z. and Mohammadi, Mohaddese and Silletta, Emilia V. and Wickenbrock, Arne and Blanchard, John W. and Budker, Dmitry and Jerschow, Alexej},
    number = {20},
    month = {5},
    pages = {10667--10672},
    volume = {117},
    publisher = {National Academy of Sciences},
    url = {https://www.pnas.org/content/early/2020/05/05/1917172117.short},
    doi = {10.1073/pnas.1917172117},
    issn = {0027-8424},
    pmid = {32376633}
}

@article{Erhard2017,
    title = {{Simulation and Measurement of the Current Density Distribution in Lithium-Ion Batteries by a Multi-Tab Cell Approach}},
    year = {2017},
    journal = {Journal of The Electrochemical Society},
    author = {Erhard, S. V. and Osswald, P. J. and Keil, P. and H{\"{o}}ffer, E. and Haug, M. and Noel, A. and Wilhelm, J. and Rieger, B. and Schmidt, K. and Kosch, S. and Kindermann, F. M. and Spingler, F. and Kloust, H. and Thoennessen, T. and Rheinfeld, A. and Jossen, A.},
    number = {1},
    month = {1},
    pages = {A6324-A6333},
    volume = {164},
    publisher = {The Electrochemical Society},
    url = {https://iopscience.iop.org/article/10.1149/2.0551701jes},
    doi = {10.1149/2.0551701jes},
    issn = {0013-4651}
}

@article{Xing2014,
    title = {{State of charge estimation of lithium-ion batteries using the open-circuit voltage at various ambient temperatures}},
    year = {2014},
    journal = {Applied Energy},
    author = {Xing, Yinjiao and He, Wei and Pecht, Michael and Tsui, Kwok Leung},
    month = {1},
    pages = {106--115},
    volume = {113},
    publisher = {Elsevier},
    url = {https://www.sciencedirect.com/science/article/pii/S0306261913005746},
    doi = {10.1016/j.apenergy.2013.07.008},
    issn = {03062619}
}

@article{Meyer2013,
    title = {{Study of the local SOC distribution in a lithium-ion battery by physical and electrochemical modeling and simulation}},
    year = {2013},
    journal = {Applied Mathematical Modelling},
    author = {Meyer, Marco and Komsiyska, Lidiya and Lenz, Bettina and Agert, Carsten},
    number = {4},
    month = {2},
    pages = {2016--2027},
    volume = {37},
    publisher = {Elsevier},
    url = {https://www.sciencedirect.com/science/article/pii/S0307904X12002600},
    doi = {10.1016/j.apm.2012.04.029},
    issn = {0307904X}
}

@article{Fleming2019,
    title = {{The design and impact of in-situ and operando thermal sensing for smart energy storage}},
    year = {2019},
    journal = {Journal of Energy Storage},
    author = {Fleming, Joe and Amietszajew, Tazdin and Charmet, Jerome and Roberts, Alexander John and Greenwood, David and Bhagat, Rohit},
    month = {4},
    pages = {36--43},
    volume = {22},
    publisher = {Elsevier},
    url = {https://www.sciencedirect.com/science/article/pii/S2352152X18306376?via%3Dihub#fig0035},
    doi = {10.1016/j.est.2019.01.026},
    issn = {2352152X}
}

@article{Murashko2013,
    title = {{Three-dimensional thermal model of a lithium ion battery for hybrid mobile working machines: Determination of the model parameters in a pouch cell}},
    year = {2013},
    journal = {IEEE Transactions on Energy Conversion},
    author = {Murashko, Kirill and Pyrhonen, Juha and Laurila, Lasse},
    number = {2},
    month = {6},
    pages = {335--343},
    volume = {28},
    url = {http://ieeexplore.ieee.org/document/6509440/},
    doi = {10.1109/TEC.2013.2255291},
    issn = {08858969}
}

@article{Xiong2018a,
    title = {{Towards a smarter battery management system: A critical review on battery state of health monitoring methods}},
    year = {2018},
    journal = {Journal of Power Sources},
    author = {Xiong, Rui and Li, Linlin and Tian, Jinpeng},
    month = {11},
    pages = {18--29},
    volume = {405},
    publisher = {Elsevier},
    url = {https://www.sciencedirect.com/science/article/pii/S037877531831111X?casa_token=i480vc215FQAAAAA:Q-OaN9cSbqR_2wihl92FoELg7jyl_yDUVYo4elEMaip32bKhUI5STw48iIMlGNU48f8p2d6_sA},
    doi = {10.1016/j.jpowsour.2018.10.019},
    issn = {03787753}
}

@article{Roth1989,
    title = {{Using a magnetometer to image a two-dimensional current distribution}},
    year = {1989},
    journal = {Journal of Applied Physics},
    author = {Roth, Bradley J. and Sepulveda, Nestor G. and Wikswo, John P.},
    number = {1},
    month = {1},
    pages = {361--372},
    volume = {65},
    publisher = {American Institute of Physics},
    url = {http://aip.scitation.org/doi/10.1063/1.342549},
    doi = {10.1063/1.342549},
    issn = {00218979}
}
\end{document}